\documentclass[journal]{IEEEtran}
\ifCLASSINFOpdf
\else
\fi
\usepackage{amsmath}
\usepackage{amssymb}
\usepackage{graphicx}
\usepackage[colorlinks=true, allcolors=blue]{hyperref}
\usepackage{tikz}
\allowdisplaybreaks[4]
\newcommand*{\circled}[1]{\lower.7ex\hbox{\tikz\draw (0pt, 0pt)%
    circle (.5em) node {\makebox[1em][c]{\small #1}};}}

\newtheorem{remark}{Remark}
\newcommand{\vg}{\boldsymbol{g}}
\newcommand{\vd}{\boldsymbol{d}}
\newcommand{\vF}{\boldsymbol{F}}

\newcommand{\mI}{\mathbf{I}}

\newcommand{\mG}{\mathbf{G}}

\newcommand{\mJ}{\mathbf{J}}
\newcommand{\vp}{\boldsymbol{p}}
\newcommand{\vv}{\boldsymbol{v}}
\newcommand{\vb}{\boldsymbol{b}}
\newcommand{\vo}{\boldsymbol{\omega}}
\newcommand{\vt}{\boldsymbol{\tau}}
\newcommand{\ve}{\boldsymbol{e}}
\newcommand{\mR}{\mathbf{R}}

\begin{document}
%
\title{Reinterpreting PID Controller From the Perspective of State Feedback and Lumped Disturbance Compensation}
%
%
%

\author{Xinyu Shi
\thanks{e-mail: {\tt yuyu.shi@qq.com}.}
}



\maketitle

\begin{abstract}
This paper analyzes the motion of solutions to non-homogeneous linear differential equations. It further clarifies that a proportional-integral-derivative (PID) controller essentially comprises two parts: a homogeneous controller and a disturbance observer, which are responsible for stabilizing the homogeneous system and compensating for the lumped disturbances (non-homogeneous components) of the system respectively. Based on this framework, the impact of measurement noise on control performance is examined, and a parameter tuning scheme for the traditional PID controller is provided. Finally, as examples, controllers are designed for two representative control problems: a trajectory tracking controller for an underactuated vertical takeoff and landing (VTOL) aircraft in the time domain, and a lateral controller for a vehicle in the distance domain.
\end{abstract}


%
\IEEEpeerreviewmaketitle

\section{Introduction}
Since its birth in the early 20th century, PID controller \cite{pid1} has been widely used in various engineering fields \cite{1453566} due to its simple structure, robustness and model-free characteristics. With the rapid development of linear algebra, increasing emphasis on Lyapunov stability analysis, and the rise of optimal control theory, various model-based control schemes—including adaptive control, sliding mode control, and optimal control methods such as linear quadratic regulator (LQR) and model predictive control (MPC)—have continually been developed. Despite the steady stream of advancements in control theory over the past half-century, PID control remains, to this day, the most widely used approach in engineering applications. \cite{Joseph2022} Aside from the widespread acceptance and reputation of PID control in engineering applications, the fundamental reason behind this phenomenon lies in the fact that the majority of control schemes are essentially linear or nonlinear combinations of system states (or state errors) of various orders, fundamentally remaining within the realm of PID control. Furthermore, these schemes often rely heavily on precise modeling, have a greater number of tuning parameters, or require more computational power. Additionally, many new control algorithms are structurally complex, making it challenging for engineering practitioners to thoroughly understand their principles and apply them effectively in real production tasks. In contrast, a simple PID control remains more comprehensible and easier to implement in today's engineering practices.

PID control requires a set of effective parameters to enable the entire system to achieve the desired performance. Consequently, empirical methods for parameter tuning, such as the Ziegler-Nichols method \cite{znopen,znclose} and relay tuning method \cite{ASTROM1984645}, have been developed. However, implementing these empirical approaches can be inefficient and complex, and the tuned parameters may still leave room for optimization. Additionally, parameter tuning methods based on linear system models have been designed \cite{772161,Borase2021,SOMEFUN202165}. Yet, due to model uncertainties, inaccuracies, unknown dynamics, and disturbances, it is challenging to analyze the system accurately in the frequency domain. As a result, these methods have not gained sufficient attention or widespread application in PID control practices in engineering.

In classical control theory, a tracking ability of system for different commands is qualitatively determined by ``type'' of the open-loop system. Influenced by this mindset, the integral component of PID controller is often interpreted individually and qualitatively as canceling steady-state error. However, is this truly the case, especially during parameter tuning? Inspired by the active disturbance rejection control (ADRC) paradigm \cite{4796887}, this paper reinterprets PID control from the perspective of state feedback for the error system and compensation for lumped disturbances, and presents a tuning method based on this approach.  The aim is to provide engineering control practitioners with a new perspective for understanding, designing, and tuning controllers.

This paper is organized as follows. Section 2 analyzes the motion of the solution to a non-homogeneous linear differential equation and presents the relationship between the final range of motion of the solution and the parameters. Section 3 reinterprets PID control from the perspective of state feedback and lumped disturbance compensation, providing a parameter tuning scheme. Section 4 illustrates the PID controller design process using two common systems as examples. Finally, Section 5 presents a summary of the paper.

\section{Motion of the Solution to a Non-Homogeneous Linear Differential Equation}

To make the analysis more general, consider a non-homogeneous linear differential equation as follows:
\begin{equation}
    x^{(n)}+\sum_{i=0}^{n-1}a_ix^{(i)}=f\label{ODE}
\end{equation}
Here, $x$ is the state variable, $\bullet^{(n)}=\frac{\mathrm{d}^n\bullet}{\mathrm{d}t^n}$, and $f=f(x,\cdots,x^{(n-1)},t)$ represents the unknown dynamics. The constants $a_i,i=0,1,\cdots, n-1$ satisfy $s^n+\sum_{i=0}^{n-1}a_is^{i}=\prod_{i=1}^n(s+p_i)$, where this is a Hurwitz polynomial with real parts of  $\mathrm{Re}(p_i)>0$. Taking the Laplace transform of the above equation under zero initial conditions yields:
\begin{align}
    X(s)=\frac{1}{s^n+\sum_{i=0}^{n-1}a_is^{i}}F(s)
\end{align}
where $X(s)=\mathcal{L}[x(t)],F(s)=\mathcal{L}[f(x(t),t)]$, $\mathcal{L}$ denotes the Laplace transform. Clearly, when the unknown dynamics $f$ are bounded, the state $x$ is also bounded.

To make the analysis more concise and clear, we now provide the solution for a special case. Consider the characteristic polynomial of the system having a repeated root, i.e. $p_i=\omega\in \mathbb{R}^+$. The differential equation can then be rewritten as
\begin{align}
    x^{(n)}=-\sum_{i=1}^nC_n^i\omega^ix^{(n-i)}+f
\end{align}
where $C_n^i=\frac{n!}{(n-i)!i!}$. In this case, the solution to the equation is
\begin{align}
x={}&x_n+x_f
\end{align}
where $x_n$  is the general solution of the homogeneous differential equation ($f=0$, corresponding to the  zero-input response of the system), and $x_f$ is a particular solution of the non-homogeneous equation with zero initial conditions (corresponding to the zero-state response of the system). Their specific forms are as follows: 
\begin{align}
    x_n={}&e^{-\omega (t-t_0)}\sum_{i=0}^{n-1}\frac{t^i}{i!}\sum_{j=0}^iC_i^j\omega^{i-j}x^{(j)}(0)\\
    x_f={}&\int_{t_0}^t\frac{(t-\tau)^{n-1}}{(n-1)!}f(\tau)e^{-\omega(t-\tau)}\mathrm{d}\tau
\end{align}
For simplicity, we have omitted the intermediate variables, i.e., $f(t)=f(x(t),\cdots,x^{(n-1)}(t),t)$. Clearly, the zero-input response of the system is always convergent, and the final motion state of the solution depends solely on the zero-state response $x_f$. Scaling the zero-state particular solution of the system gives:
\begin{align}
    |x_f|={}&\left|\int_{t_0}^t\frac{(t-\tau)^{n-1}}{(n-1)!}f(\tau)e^{-\omega(t-\tau)}\mathrm{d}\tau\right|\notag\\
\leq{}&\int_{t_0}^t\frac{(t-\tau)^{n-1}}{(n-1)!}|f(\tau)|e^{-\omega(t-\tau)}\mathrm{d}\tau\notag\\
\leq{}&\int_{t_0}^t\frac{(t-\tau)^{n-1}}{(n-1)!}\sup_{s\geq\tau}|f(s)|e^{-\omega(t-\tau)}\mathrm{d}\tau
\end{align}
Taking the limit of the above equation yields:
\begin{align}
    \lim_{t\to+\infty}|x_f|\leq{}&\lim_{t\to+\infty}\frac{\int_{t_0}^t\frac{(t-\tau)^{n-1}}{(n-1)!}\sup_{s\geq\tau}|f(s)|e^{\omega(\tau-t_0)}\mathrm{d}\tau}{e^{\omega (t-t_0)}}\notag\\
    ={}&\lim_{t\to+\infty}\frac{\int_{t_0}^t\frac{(t-\tau)^{n-2}}{(n-2)!}\sup_{s\geq\tau}|f(s)|e^{\omega (\tau-t_0)}\mathrm{d}\tau}{\omega e^{\omega (t-t_0)}}\notag\\
    \vdots{}&\notag\\
    ={}&\lim_{t\to+\infty}\frac{\int_{t_0}^t\sup_{s\geq\tau}|f(s)|e^{\omega (\tau-t_0)}\mathrm{d}\tau}{\omega^{n-1} e^{\omega (t-t_0)}}\notag\\
    ={}&\lim_{t\to+\infty}\frac{\sup_{s\geq t}|f(s)|e^{\omega (t-t_0)}}{\omega^n e^{\omega (t-t_0)}}\notag\\
    ={}&\frac{1}{\omega^n}\limsup_{t\to+\infty}|f(t)|
\end{align}
Therefore, a sufficient condition for the asymptotic stability of the zero equilibrium of the system is $\limsup_{t\to+\infty}|f(t)|=0$. It is also evident that when $\bar f:=\limsup_{t\to+\infty}|f(t)|$  is a small constant, the final absolute value of the state of the system is very small. The transfer function of the system is:
\begin{align}
    G(s)=\frac{X(s)}{F(s)}=\frac{1}{(s+\omega)^n}
\end{align}
It can be seen that the state variable $x$ is the $n$-th order low-pass filtered output of $f$, with an output gain of $\omega^{-n}$. Therefore, the output depends on two factors:
\begin{enumerate}
\item The bandwidth $\omega$: the larger the bandwidth, the smaller the gain of the filter, and thus the smaller the output.
\item The unknown quantity $f$: the smaller $f$ is, the smaller the output will be.
\end{enumerate}
The bandwidth is constrained by the physical properties of the plant and typically has an upper limit. To minimize the final value of the solution, the unknown dynamics $f$  should be made as small as possible.

\section{State Feedback and Lumped Disturbance Compensation}
From the analysis in the previous section, the following conclusion is readily apparent: for a bounded-input bounded-output (BIBO) system (\ref{ODE}) with unknown dynamics, the ultimate convergence behavior depends on the unknown dynamics $f$. Therefore, in this section, we propose a scheme to compensate for the system’s unknown dynamics to achieve better control performance.
\subsection{Control Design and Tuning Scheme}
For the state $x$, control input $u$, unknown lumped disturbance $f=f(x,\cdots,x^{(n-1)},u,t)$ , and input coefficient $b\neq 0$, the non-autonomous system is as follows:
\begin{align}
    x^{(n)} = f+bu,n\in\mathbb{N}^+\label{system}
\end{align}
The internal model principle reveals that the key to achieving good control performance lies in compensating for the unknown lumped disturbance $f$. Here, $f$  includes uncertain signals such as unknown dynamics and disturbances. Clearly, it is nearly impossible to observe $f$ without error; therefore, a reasonably simple observer must be designed to appropriately compensate for the lumped disturbance. At the same time, we need to stabilize the homogeneous system ($f=0$). For this purpose, we divide the control input $u$ into two components, as follows:
\begin{align}
    u=\frac{1}{b}(u_x-\hat{f})\label{input}
\end{align}
where $\hat{f}$ is used to estimate (observe) and compensate for the unknown lumped disturbance $f$ and $u_x$, referred to as the homogeneous controller, is used to drive the entire homogeneous system to convergence. Based on the analysis in the previous section, design
\begin{align}
    u_x=-\sum_{i=0}^{n-1}a_ix^{(i)}
\end{align}
and define the observation error of the unknown dynamics as
\begin{align}
    \tilde{f}:=f-\hat{f}\label{obserror}
\end{align}
Substituting Eqs. (\ref{input}) and (\ref{obserror}) into the system (\ref{system}) yields:
\begin{align}
    x^{(n)}-u_x=\tilde{f}\label{systemtrans}
\end{align}
The observation component is designed as follows:
\begin{align}
    \hat f = \omega_f\left(x^{(n-1)} -  \int_0^tu_x\mathrm{d}t\right)\label{observer}
\end{align}
Substituting the system model (\ref{systemtrans}) into the above equation and performing an identity transformation yields:
\begin{align}
    \hat f=\omega_f \int_0^t\tilde{f}\mathrm{d}t=\omega_f \int_0^t\left(f-\hat{f}\right)\mathrm{d}t
\end{align}
\begin{remark}
For simplicity, the term $\omega_fx^{(n-1)}(0)$ is omitted here. Since it can be considered part of the unknown lumped disturbance $f$, this omission does not affect the conclusions. The rationale for omitting initial values of integrals in subsequent analysis in this section is consistent with this explanation and will not be repeated.
\end{remark}

Taking the Laplace transform of the above equation under zero initial conditions, the transfer function of this observer is given by:
\begin{align}
    G_o(s)=\frac{\hat{F}(s)}{F(s)}=\frac{\omega_f}{s+\omega_f}
\end{align}
Further, the observation error is given by:
\begin{align}
    \tilde{f}=f-\omega_f\int_0^t\tilde{f}\mathrm{d}t
\end{align}
It can also be seen that the error transfer function is:
\begin{align}
    G_e(s)=\frac{\tilde{F}(s)}{F(s)}=\frac{s}{s+\omega_f}
\end{align}
It can be seen that the observer bandwidth $\omega_f$ affects the performance of the observer. The larger the bandwidth, the closer the transfer function of the observer is to 1, resulting in a smaller observation error, and vice versa. Furthermore, it follows that the transfer function of the entire system is:
\begin{align}
    G(s)=\frac{X(s)}{F(s)}=\frac{s}{(s^n+\sum_{i=0}^{n-1}a_is^{i})(s+\omega_f)}
\end{align}
The impact of noise on control performance cannot be ignored. The effect of noise amplification due to the gains needs to be qualitatively analyzed in order to find the optimal control parameters. In engineering practice, it is nearly impossible to measure each state variable without error; therefore, the measured values are defined as follows:
\begin{align}
    z_i = x^{(i)}+w_i,i=0,1,\cdots, n-1
\end{align}
where $w_i$ represents the measurement noise. By replacing $x_i$ with $z_i$ in the controller, the actual control input is computed as follows:
\begin{align}
    u=\frac{1}{b}(u_x+u_w-\hat{f}-\hat{w})\label{inputwithnoise}
\end{align}
with
\begin{align}
    u_w&=-\sum_{i=0}^{n-1}a_iw_{n-i},\\
    \hat{w}&=\omega_f\left(w_{n-1} - \int_0^tu_w\mathrm{d}t\right)\approx \omega_f w_{n-1}
\end{align}
Substituting the control signal (\ref{inputwithnoise}) that considers measurement noise into system (\ref{system}) and performing the Laplace transform yields:
\begin{align}
   X(s)={}&\frac{s(F(s)+U_x(s)-\hat{W}(s)}{(s^n+\sum_{i=0}^{n-1}a_is^{i})(s+\omega_f)}\notag\\
   ={}&\frac{\omega_fU_w(s)}{(s^n+\sum_{i=0}^{n-1}a_is^{i})(s+\omega_f)}\notag\\
   &+\frac{s(F(s)+U_w(s)-\omega_fW_{n-1})}{(s^n+\sum_{i=0}^{n-1}a_is^{i})(s+\omega_f)}\notag\\
   \approx{}&\frac{s(F(s)+U_w(s)-\omega_fW_{n-1})}{(s^n+\sum_{i=0}^{n-1}a_is^{i})(s+\omega_f)}
\end{align}
From the above equation, it can be seen that part of the high-frequency measurement noise in $\hat{w}$ is filtered through integration and low-pass filtering, while the remaining portion, along with the noise $u_w$ from the homogeneous controller, affects the entire closed-loop system. Therefore, during parameter tuning, the bandwidth should be carefully selected based on the noise frequency. A reasonable approach is to set the homogeneous controller bandwidth smaller than the observer bandwidth to minimize the impact of noise on the system performance. This is similar to the separation principle, where the parameters of the observer and the homogeneous controller can be designed separately without considering their coupling. We can simplistically assume that the reference of the homogeneous controller only determines the system’s poles, while the observer parameters only determine the observer’s bandwidth. In this way, the loop shaping process of a system is decomposed into tuning two relatively low-order closed-loop systems. However, for the homogeneous controller, there are still multiple parameters that need to be tuned. This paper recommends configuring the poles of the homogeneous system to be the same value \cite{1242516}, that is :
\begin{align}
    u_x=-\sum_{i=1}^nC_n^i\omega^ix^{(n-i)}
\end{align}
The benefits of this approach are evident: the entire homogeneous controller has only one parameter, $\omega$. Its physical significance, along with its reciprocal $T=\frac{1}{\omega}$, is also clear, representing the bandwidth and the time constant of the dynamic response. In addition, to achieve better performance from the PID controller, it is best to know the input coefficient $b$ of the controller, even if only an approximate value.

\subsection{PI and PID Controllers}
In the above analysis, we have essentially examined the functions of the various components of the ``generalized PID'' controller, namely, the state feedback component for stabilizing the homogeneous system and the observation component for compensating lumped disturbances. The reason for using ``generalized PID'' to describe the above controller is that it is merely a higher-order extension and generalization of the PID controller for second-order systems. By applying simple simplifications, one can derive the PI controller for first-order systems and the PID controller for second-order systems. To make the description clearer and in comparison with the PID, we will briefly outline this process below.
First, we simplify the above system (\ref{system}) to a first-order system, which gives:
\begin{align}
    \dot{x}=  f+bu
\end{align}
Accordingly, the homogeneous controller and the lumped disturbance observer are:
\begin{align}
    u_x={}&-a_0 x\\
    \hat f ={}& -\omega_f\int_0^tu_x\mathrm{d}t
\end{align}
Substituting $u_x$ and $\hat{f}$ into the controller $u$, we get:
\begin{align}
u={}&\frac{1}{b}\left(-a_0 x-\omega_fa_0\int_0^t x\mathrm{d}t\right)
\end{align}
Clearly, the above equation represents a PI controller ($k_p=a_0, k_i=\omega_fa_0$).

Similarly, when system (\ref{system}) is instantiated as a second-order system, we have:

\begin{align}
    \ddot{x}=  f+bu
\end{align}
Accordingly, the homogeneous controller and the lumped disturbance observer are:
\begin{align}
    u_x={}&-a_1\dot{x}-a_0 x\\
    \hat f ={}& \omega_f\left(\dot{x} -  \int_0^tu_x\mathrm{d}t\right)
\end{align}
Substituting $u_x$ and $\hat{f}$ into the controller $u$, we get:
\begin{align}
u={}&\frac{1}{b}\left(-a_1\dot{x}-a_0 x-\omega_f\left(\dot{x}- \int_0^t (-a_1\dot{x}-a_0 x)\mathrm{d}t\right)\right)\notag\\
={}&\frac{1}{b}\left(-(a_1+\omega_f)\dot{x}-(a_0+\omega_fa_1) x-\omega_fa_0\int_0^tx\mathrm{d}t\right)
\end{align}
Clearly, the above equation represents a PID controller ($k_d=a_1+\omega_f,k_p=a_0+\omega_fa_1,k_i=\omega_fa_0$).

\section{Control Design Examples}
In this section, we will design controllers for two representative physical systems.
\subsection{Trajectory Tracking Control for Underactuated VTOL Aircraft}

The underactuated VTOL aircraft, as a 6-degree-of-freedom (DOF) rigid body, can be used to describe the models of quadrotors, helicopters, and other aircraft. Its mathematical model is as follows:
\begin{align}
    \left\{\begin{aligned}\dot{\vp}={}&\vv\\
    m\dot{\vv}={}&mg\ve_3-f\mR \ve_3+\vd_f\\
    \dot{\mR}={}&\mR \vo_\times\\
    \mJ\dot{\vo}={}&-\vo\times (\mJ\vo)+\vt+\vd_\tau\end{aligned}\right.
\end{align}
where $\vp,\vv,\vo\in\mathbb{R}^3$ represent position, velocity, and angular velocity, respectively; $\mR\in\mathrm{SO}(3)$ describes the attitude of the rigid body; $m$ is the mass; $g$ is the gravitational acceleration; and $\mJ\in\mathbb{R}^{3\times 3}$ is the inertia matrix. $\vd_f,\vd_\tau\in\mathbb{R}^3$ are unknown disturbance signals, $f\in \mathrm{R}$ is the thrust in this subsection, $\vt\in\mathbb{R}^3$ is the torque, and $\ve_3=[0,0,1]^\top$. The subscript $\times$ describes the following mapping:
\begin{align}
    \left[\begin{matrix}
        x_1\\x_2\\x_3
    \end{matrix}\right]_\times=\left[\begin{matrix}
        0&-x_3&x_2\\
        x_3&0&-x_1\\
        -x_2&x_1&0
    \end{matrix}\right]
\end{align}
The control objective is to track the 3D trajectory $\vp_d\in\mathbb{R}^3$ and the flight orientation angle $\psi_d$. By simplifying our previous work \cite{Sujie2024}, we obtain the error model:
\begin{align}
    \left\{\begin{aligned}\ddot{\tilde{\vp}}={}&
    g\ve_3-\frac{f}{m}\mR\ve_3+\frac{1}{m}\vd_f-\ddot{\vp}_d\\
    \ddot{\tilde{\vg}}={}&\mG(\mJ^{-1}(-\vo_\times \mJ\vo+\vt+\vd_\tau)+(\mG^{-1}\dot{\tilde{\vg}})_\times\vo-\tilde{\mR}^\top\dot{\vo}_d)\\
    &+\frac{2\tilde{\vg}^\top\dot{\tilde{\vg}}\dot{\tilde{\vg}}}{1+\tilde{\vg}^\top\tilde{\vg}}\end{aligned}\right.\label{vtolerrormodel}
\end{align}
where $\mR_d=[\vb_{1d},\vb_{2d}, \vb_{3d}]=[\vb_{2d\times}\vb_{3d},\frac{\vb_{3d\times}\vb_d}{\Vert\vb_{3d\times}\vb_d\Vert},\frac{\vF_d}{\Vert \vF_d\Vert}]\in\mathrm{SO}(3)$ describes the desired attitude with the desired control thrust $\vF_d\in\mathbb{R}^3$. $\vp_d\in\mathbb{R}^3$denotes reference position. $\tilde{\vp}=\vp-\vp_d,\tilde{\mR}=\mR_d^\top\mR$ represent the position and attitude errors, respectively.  The angular velocity error is described by $\tilde{\vo}=\vo-\tilde{\mR}^\top\vo_d$, with the desired attitude angular velocity given by $\vo_{d\times}={\mR}_d^\top\dot{\mR}_d$. The calculations for the remaining relevant variables are as follows:
\begin{align}   
    \vb={}&[\cos\psi_d,\sin\psi_d,0]^\top,\notag\\
    \tilde{\vg}={}&\frac{(\tilde{\mR}-\tilde{\mR}^\top)^\times}{\mathrm{tr}(\tilde{\mR})+1},\notag\\
    \mG={}&\frac{1}{2}(\mI+\tilde{\vg}_\times+\tilde{\vg}\tilde{\vg}^\top),\notag\\
    \dot{\tilde{\vg}}={}&\mG\tilde{\vo}\notag
\end{align}
The control force is extracted using a projection method as follows:
\begin{align}
    f=\ve_3^\top\mR^\top\vF_d
\end{align}
The error system model (\ref{vtolerrormodel}) is divided into input signals and lumped disturbances, as follows:
\begin{align}
    \left\{\begin{aligned}\ddot{\tilde{\vp}}={}&
    \frac{1}{m}\vF+\vd_f^{\mathrm{lump}}\\
    \ddot{\tilde{\vg}}={}&\mG\mJ^{-1}\vt+\vd_\tau^{\mathrm{lump}}\end{aligned}\right.
\end{align}
where
\begin{align}
    \vF={} &mg\ve_3-f\mR\ve_3,\vd_f^{\mathrm{lump}}=\frac{1}{m}\vd_f-\ddot{\vp}_d\notag\\
\vd_\tau^{\mathrm{lump}}={} &\mG(\mJ^{-1}(-\vo_\times (\mJ\vo)+\vd_\tau)+(\mG^{-1}\dot{\tilde{\vg}})_\times\vo-\tilde{\mR}^\top\dot{\vo}_d)\notag\\
&+\frac{2\tilde{\vg}^\top\dot{\tilde{\vg}}\dot{\tilde{\vg}}}{1+\tilde{\vg}^\top\tilde{\vg}}\notag
\end{align}
The controller is designed as follows:
\begin{align}
    \left\{\begin{aligned}\vF_d&=-m(\vF_x-\hat{\vd}_f^\mathrm{lump}-g\ve_3)\\
    \vF_x&=-k_0\tilde{\vp}-k_1\dot{\tilde{\vp}}\\
    \hat{\vd}_f^\mathrm{lump}&=\omega_f\left(\dot{\tilde{\vp}}-\int_0^t\vF_x\mathrm{d}t\right)\\
    \vt&=\mJ\mG^{-1}(\vt-\hat{\vd}_\tau^\mathrm{lump})\\
\vt_x&=-k_0\tilde{\vg}-k_1\dot{\tilde{\vg}}\\
    \hat{\vd}_\tau^\mathrm{lump}&=\omega_\tau\left(\dot{\tilde{\vg}}-\int_0^t\vt_x\mathrm{d}t\right)\end{aligned}\right.
\end{align}

\subsection{Lateral Control of the Vehicle}\label{sebsectionLCV}
The kinematic model of the vehicle is typically described by the bicycle model as follows:

\begin{align}
    \left\{\begin{aligned}\dot{x}={}&v\cos\theta\\
    \dot{y}={}&v\sin\theta\\
    \dot{\theta}={}&\frac{v\tan(\delta+d)}{L}\end{aligned}\right.
\end{align}
where $x,y\in\mathbb{R}$ represent the coordinates of the rear axle center of the vehicle, $\theta$ is the yaw angle, $v\in\mathbb{R}$ is the speed scalar at the midpoint of the rear axle, $L$ is the wheelbase, $\delta$ is the steering angle controlled by the steering wheel, and $d$ represents a matching disturbance that can describe the bias of the steering mechanism.
\begin{figure}[htp]
  \centering
  \includegraphics[height=4.0cm]{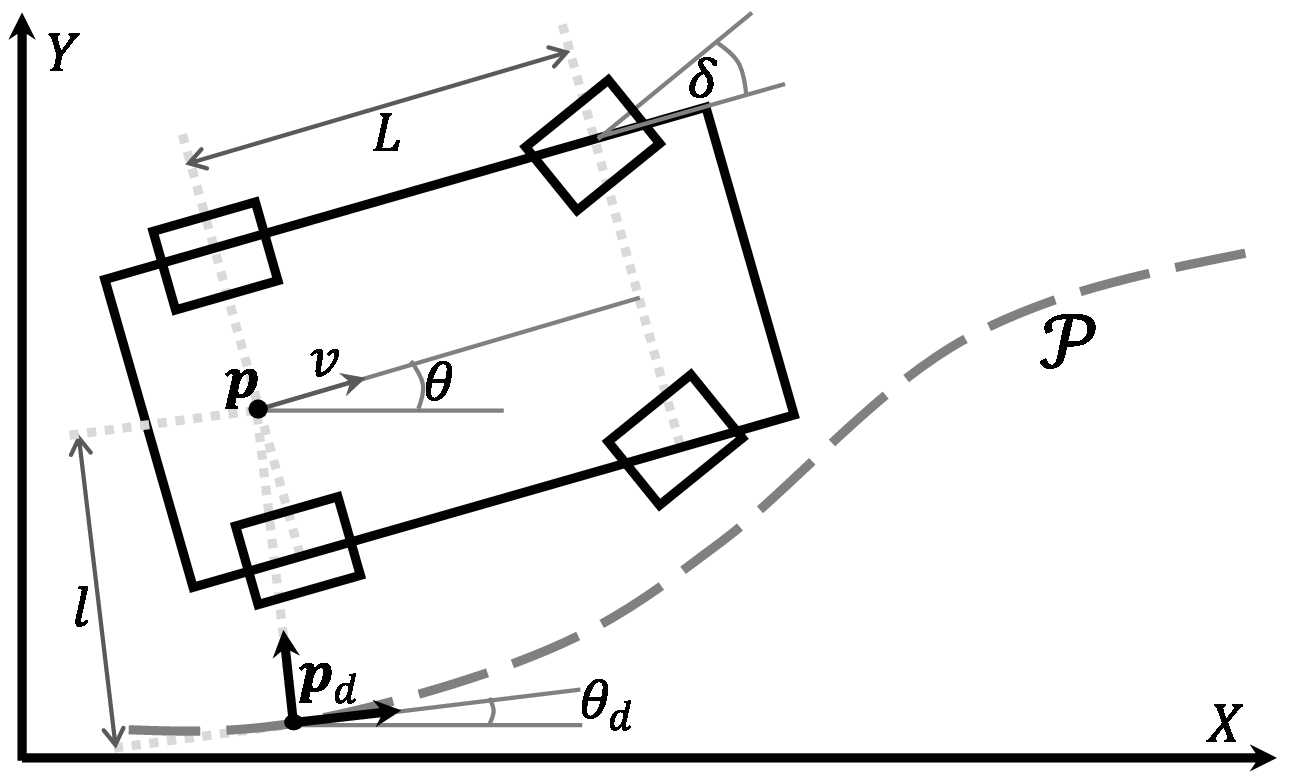}
  \caption{Schematic diagram of vehicle lateral path tracking in Frenet--Serret frame}
  \label{figlattrack}
\end{figure} 
For lateral control in path tracking problems, we do not focus on the speed of the vehicle information; therefore, we can convert the kinematics to the distance (arc-length) domain, as follows:
\begin{align}
    \left\{\begin{aligned}x'={}&\cos\theta\\
    y'={}&\sin\theta\\
    \theta'={}&\frac{\tan(\delta+d)}{L}\end{aligned}\right.
\end{align}
Where $\bullet'=\frac{\mathrm{d}\bullet}{\mathrm{d}s}$, and in this section, $s$ represents the distance traveled by the vehicle, with $\dot{s}=v$.
As As shown in Figure \ref{figlattrack}, for a given reference path $\mathcal{P}:\mathbb{R}^+\to\mathbb{R}^2\times\mathbb{S}$, $\vp_d=[x_d,y_d,\theta_d]^\top$ is the matching point of the vehicle on $\mathcal{P}$ in the Frenet-Serret frame \cite{Frenet1852,Serret1851}. The error is defined as:
\begin{align}
e_x=x_d-x,e_y=y_d-y,e_\theta=\theta_d-\theta
\end{align}
According to the definition of the Frenet-Serret frame, the following constraints hold:
\begin{align}
    \left[\begin{matrix}
        0\\
        l
    \end{matrix}\right]=\left[\begin{matrix}
        \cos\theta_d&\sin\theta_d\\
        -\sin\theta_d&\cos\theta_d
    \end{matrix}\right]\left[\begin{matrix}
        e_x\\
        e_y
    \end{matrix}\right]=
    \left[\begin{matrix}
        e_x\cos\theta_d+e_y\sin\theta_d\\
        -e_x\sin\theta_d+e_y\cos\theta_d
    \end{matrix}\right]\label{latconstrains}
\end{align}
where $l$ represents the lateral error. To facilitate the subsequent development of the error model, we assume that $e_\theta\in(-\frac{\pi}{2},\frac{\pi}{2})$, and that for the path arc length $s_d$, $r_s:=\frac{\mathrm{d}s_d}{\mathrm{d}s}$ exists. In fact, for $v\neq 0$ we have $r_s=\frac{v_d}{v}$, where $v_d=\sqrt{\dot{x}_d^2+\dot{y}_d^2}$ is the velocity of the matching point, rather than the result of the velocity planning. Additionally, it can be noted that if the lateral control tracking error is very small, we can assume that $r_s\approx 1$. By substituting the above constraint equations into the derivative of $l$ with respect to $s$, we get:
\begin{align}
    l'={}&-r_s\kappa_d(e_x\cos\theta_d +e_y\sin\theta_d)-\sin\theta_d(r_s\cos\theta_d-\cos\theta)\notag\\
    &+\cos\theta_d(r_s\sin\theta_d-\sin\theta)\\
={}&\sin(\theta_d-\theta)=\sin e_\theta \label{latfirstorder}
\end{align}
where
\begin{align}
    \left\{\begin{aligned}\frac{\mathrm{d}x_d}{\mathrm{d}s_d}={}&\cos\theta_d\\
   \frac{\mathrm{d}y_d}{\mathrm{d}s_d}={}&\sin\theta_d\\
    \frac{\mathrm{d}\theta_d}{\mathrm{d}s_d}={}&\kappa_d\end{aligned}\right. \Longrightarrow\  \left\{\begin{aligned}\frac{\mathrm{d}x_d}{\mathrm{d}s}={}&r_s\cos\theta_d\\
   \frac{\mathrm{d}y_d}{\mathrm{d}s}={}&r_s\sin\theta_d\\
    \frac{\mathrm{d}\theta_d}{\mathrm{d}s}={}&r_s\kappa_d\end{aligned}\right.
\end{align}
Continuing with the differentiation, we get:
\begin{align}
    l''={}&\cos e_\theta(r_s\kappa_d-\frac{\tan(\delta+d)}{L})\label{latsecondorder}
\end{align}
where $\bullet''=\frac{\mathrm{d}^2\bullet}{\mathrm{d}s^2}$. If $d$ is known and assuming $r_s=1$, we can directly design the state feedback control law as follows:
\begin{align}
    \delta=\arctan\left(L(\kappa_d+\sec e_\theta(k_0l+k_1\sin e_\theta))\right)-d
\end{align}
\begin{remark}  As shown in Figure \ref{figlattrackvehicle}, if the intersection of path $\mathcal{P}$ and the rear axle of the vehicle is taken as the projection point $\vp_d$, the lateral error is still defined as the distance between $\vp$ and $\vp_d$. It is only necessary to transform the errors $e_x$ and $e_y$ into the vehicle coordinate system to obtain constraints similar to Eq. (\ref{latconstrains}). By taking the first and second derivatives of $l$ with respect to $s$, the model for the lateral error can be derived. The specific process is not elaborated further in this paper. It is worth noting that although the selection scheme for matching may differ, the lateral error model is exactly the same as the one in equations (\ref{latfirstorder}) and (\ref{latsecondorder}).
\begin{figure}[htp]
  \centering
  \includegraphics[height=4.0cm]{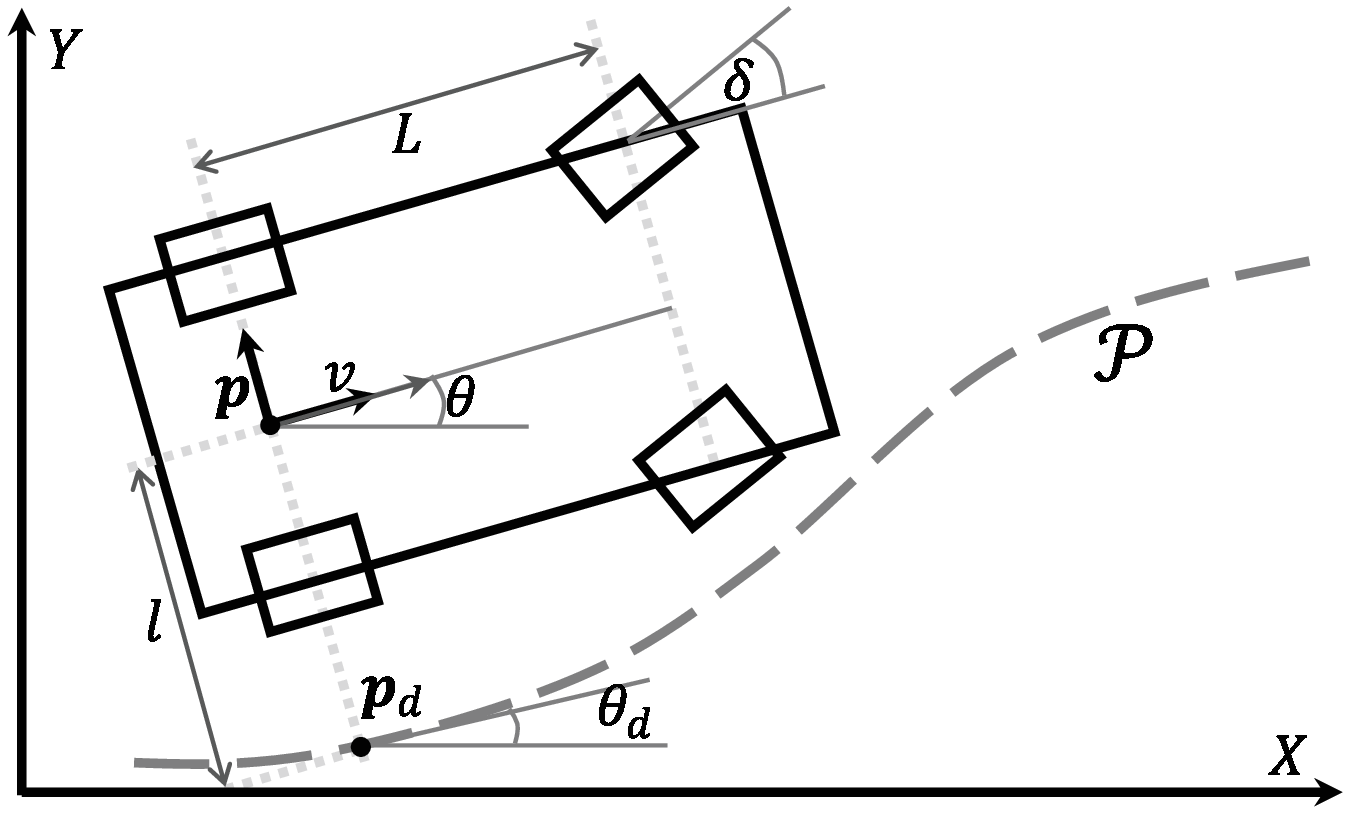}
  \caption{Schematic diagram of vehicle lateral path tracking in vehicle frame}
  \label{figlattrackvehicle}
\end{figure} 
\end{remark}
When the disturbance is unknown, the error model is transformed as follows:
\begin{align}
    l''={}&-\frac{\cos e_\theta\tan\delta}{L}+d^{\mathrm{lump}}\\
    d^{\mathrm{lump}}={}&\cos e_\theta\left(r_s\kappa_d-\frac{\tan d(1+\tan^2\delta)}{L(1-\tan\delta\tan d)}\right)\notag
\end{align}
The controller is designed as follows:
\begin{align}
\left\{\begin{aligned}\delta&=\arctan(L\sec e_\theta(u_x+\hat{d}^\mathrm{lump}))\\
    u_x&=k_0l+k_1l'=k_0l+k_1\sin e_\theta\\
        \hat{d}^\mathrm{lump}&=\omega_d(l'+\int_0^su_x\mathrm{d}s)=\omega_d(\sin e_\theta+\int_0^tvu_x\mathrm{d}t)\end{aligned}\right.
\end{align}
\section{Discussion and Conclusion}

This paper reinterprets the principle of PID controllers by drawing on the motion of the solution to non-homogeneous linear differential equations with constant coefficients. The PID controller consists of two components: the state feedback of the homogeneous system and the compensation of lumped disturbances. Based on this, we present a parameter tuning scheme for PID controllers. For the homogeneous system, appropriate pole placement is sufficient to achieve the desired convergence behavior. As for the compensation of lumped disturbances, we only need to adjust the bandwidth of the observer. Although the scheme for compensating lumped disturbances in PID is simple, it is not always the most suitable. Fortunately, we now clearly understand that the controller essentially consists of two parts: the homogeneous system controller and the lumped disturbance observer. Therefore, we can choose different schemes for observing and compensating lumped disturbances. In this context, we recommend the ADRC paradigm to engineers, as it is the main inspiration for this paper. Below, we outline the advantages of Extended State Observer (ESO \cite{4796887}, including linear ESO (LESO) \cite{1242516} and generalized ESO (GESO) \cite{6117083}) in the ADRC control paradigm:
\begin{enumerate}
\item The ESO requires minimal prior knowledge of the system, needing only the system order, input coefficients, and output signals.
\item Not all state variables of a system are always available, and numerically differentiating the output to approximate derivatives often leads to the amplification of noise. The ESO, being an output feedback-based observer, can observe both lumped disturbances and all the state variables of the system.
\item For multi-input, multi-output systems with mismatched disturbances, it is difficult to estimate the disturbances using conventional methods. However, the GESO provides a feasible solution.
\item The parameters of the ESO typically only determine the performance of the observer. We can independently assign zeros and poles to both the observer and the controller, which means that the separation principle remains applicable in most cases.
\end{enumerate}
Finally, for two common control problems—the trajectory tracking of an underactuated VTOL aircraft and the lateral control of a vehicle—we have designed PID controllers based on their error models. Here, we summarize the control design process and present a more general set of design steps:
\begin{enumerate}
    \item Deriving the Mathematical Model of the Error in the Control Problem: Ideally, the error model should be in the form of an integral cascade structure, as this simplifies both controller and observer design. For this purpose, input-output feedback linearization \cite{khalil2002nonlinear} is often a useful theoretical tool. Additionally, for certain control tasks, it may be beneficial to break free from traditional thinking and build models in alternative dimensions, as demonstrated in Section \ref{sebsectionLCV}, where the controller is designed in the distance domain.
    \item Decomposing the Error Model into Input Signals and Lumped Disturbances: It is important to separate the error system into these two components for a more structured approach to control design.
    \item Selecting a Control Scheme and Designing the Controller: If the mathematical model of the error system is simple and all relevant information is known, one mFrenetay directly design the control law without the need for integrators or observers. For more complex cases, the PID controller introduced in this paper or the ADRC paradigm can be applied.
\end{enumerate}


%



\ifCLASSOPTIONcaptionsoff
  \newpage
\fi

\bibliographystyle{IEEEtran}
\bibliography{ref}

\begin{thebibliography}{10}
\providecommand{\url}[1]{#1}
\csname url@samestyle\endcsname
\providecommand{\newblock}{\relax}
\providecommand{\bibinfo}[2]{#2}
\providecommand{\BIBentrySTDinterwordspacing}{\spaceskip=0pt\relax}
\providecommand{\BIBentryALTinterwordstretchfactor}{4}
\providecommand{\BIBentryALTinterwordspacing}{\spaceskip=\fontdimen2\font plus
\BIBentryALTinterwordstretchfactor\fontdimen3\font minus \fontdimen4\font\relax}
\providecommand{\BIBforeignlanguage}[2]{{%
\expandafter\ifx\csname l@#1\endcsname\relax
\typeout{** WARNING: IEEEtran.bst: No hyphenation pattern has been}%
\typeout{** loaded for the language `#1'. Using the pattern for}%
\typeout{** the default language instead.}%
\else
\language=\csname l@#1\endcsname
\fi
#2}}
\providecommand{\BIBdecl}{\relax}
\BIBdecl

\bibitem{pid1}
\BIBentryALTinterwordspacing
N.~Minorsky., ``Directional stability of automatically steered bodies,'' \emph{Journal of the American Society for Naval Engineers}, vol.~34, no.~2, pp. 280--309, 1922. [Online]. Available: \url{https://onlinelibrary.wiley.com/doi/abs/10.1111/j.1559-3584.1922.tb04958.x}
\BIBentrySTDinterwordspacing

\bibitem{1453566}
K.~H. Ang, G.~Chong, and Y.~Li, ``Pid control system analysis, design, and technology,'' \emph{IEEE Transactions on Control Systems Technology}, vol.~13, no.~4, pp. 559--576, 2005.

\bibitem{Joseph2022}
\BIBentryALTinterwordspacing
S.~B. Joseph, E.~G. Dada, A.~Abidemi, D.~O. Oyewola, and B.~M. Khammas, ``Metaheuristic algorithms for pid controller parameters tuning: review, approaches and open problems,'' \emph{Heliyon}, vol.~8, no.~5, p. e09399, 2022. [Online]. Available: \url{https://www.sciencedirect.com/science/article/pii/S2405844022006879}
\BIBentrySTDinterwordspacing

\bibitem{znopen}
\BIBentryALTinterwordspacing
J.~G. Ziegler and N.~B. Nichols, ``{Optimum Settings for Automatic Controllers},'' \emph{Transactions of the American Society of Mechanical Engineers}, vol.~64, no.~8, pp. 759--765, 12 1942. [Online]. Available: \url{https://doi.org/10.1115/1.4019264}
\BIBentrySTDinterwordspacing

\bibitem{znclose}
\BIBentryALTinterwordspacing
------, ``{Process Lags in Automatic-Control Circuits},'' \emph{Transactions of the American Society of Mechanical Engineers}, vol.~65, no.~5, pp. 433--440, 12 1943. [Online]. Available: \url{https://doi.org/10.1115/1.4018788}
\BIBentrySTDinterwordspacing

\bibitem{ASTROM1984645}
\BIBentryALTinterwordspacing
K.~Åström and T.~Hägglund, ``Automatic tuning of simple regulators with specifications on phase and amplitude margins,'' \emph{Automatica}, vol.~20, no.~5, pp. 645--651, 1984. [Online]. Available: \url{https://www.sciencedirect.com/science/article/pii/0005109884900141}
\BIBentrySTDinterwordspacing

\bibitem{772161}
Q.-G. Wang, T.-H. Lee, H.-W. Fung, Q.~Bi, and Y.~Zhang, ``Pid tuning for improved performance,'' \emph{IEEE Transactions on Control Systems Technology}, vol.~7, no.~4, pp. 457--465, 1999.

\bibitem{Borase2021}
\BIBentryALTinterwordspacing
R.~P. Borase, D.~K. Maghade, S.~Y. Sondkar, and S.~N. Pawar, ``A review of pid control, tuning methods and applications,'' \emph{International Journal of Dynamics and Control}, vol.~9, no.~2, pp. 818--827, Jun 2021. [Online]. Available: \url{https://doi.org/10.1007/s40435-020-00665-4}
\BIBentrySTDinterwordspacing

\bibitem{SOMEFUN202165}
\BIBentryALTinterwordspacing
O.~A. Somefun, K.~Akingbade, and F.~Dahunsi, ``The dilemma of pid tuning,'' \emph{Annual Reviews in Control}, vol.~52, pp. 65--74, 2021. [Online]. Available: \url{https://www.sciencedirect.com/science/article/pii/S1367578821000407}
\BIBentrySTDinterwordspacing

\bibitem{4796887}
J.~Han, ``From pid to active disturbance rejection control,'' \emph{IEEE Transactions on Industrial Electronics}, vol.~56, no.~3, pp. 900--906, 2009.

\bibitem{1242516}
Z.~Gao, ``Scaling and bandwidth-parameterization based controller tuning,'' in \emph{Proceedings of the 2003 American Control Conference, 2003.}, vol.~6, 2003, pp. 4989--4996.

\bibitem{Sujie2024}
\BIBentryALTinterwordspacing
S.~Zhang, X.~Shi, X.~Chen, and X.~He, ``Trajectory tracking control based on generalized rodrigues parameter for underactuated vtol uavs,'' \emph{Journal of Aerospace Engineering}, vol.~37, no.~6, p. 04024095, Nov 2024. [Online]. Available: \url{https://doi.org/10.1061/JAEEEZ.ASENG-5449}
\BIBentrySTDinterwordspacing

\bibitem{Frenet1852}
\BIBentryALTinterwordspacing
F.~Frenet, ``\BIBforeignlanguage{fre}{Sur les courbes à double courbure.}'' \emph{\BIBforeignlanguage{fre}{Journal de Mathématiques Pures et Appliquées}}, pp. 437--447, 1852. [Online]. Available: \url{http://eudml.org/doc/233946}
\BIBentrySTDinterwordspacing

\bibitem{Serret1851}
\BIBentryALTinterwordspacing
J.-A. Serret, ``\BIBforeignlanguage{fre}{Sur quelques formules relatives à la théorie des courbes à double courbure.}'' \emph{\BIBforeignlanguage{fre}{Journal de Mathématiques Pures et Appliquées}}, pp. 193--207, 1851. [Online]. Available: \url{http://eudml.org/doc/235002}
\BIBentrySTDinterwordspacing

\bibitem{6117083}
S.~Li, J.~Yang, W.-H. Chen, and X.~Chen, ``Generalized extended state observer based control for systems with mismatched uncertainties,'' \emph{IEEE Transactions on Industrial Electronics}, vol.~59, no.~12, pp. 4792--4802, 2012.

\bibitem{khalil2002nonlinear}
H.~Khalil, \emph{Nonlinear systems (Third Edition)}.\hskip 1em plus 0.5em minus 0.4em\relax Prentice Hall, 2002.

\end{thebibliography}

\end{document}